\documentclass[conference]{IEEEtran}
\usepackage{blindtext}
\usepackage{graphicx}

\ifCLASSINFOpdf
  \graphicspath{{./figures/}}
\else
\fi
\usepackage{url}

\usepackage{siunitx}

\usepackage{color}

\usepackage{tikz}
\usetikzlibrary{calc}
\newcommand{\tikzCircle}[1]{%
  \tikz[baseline={([yshift=-8pt]current bounding box.north)}] \path (0,0) node[inner sep=0.5pt,minimum size=2mm,draw,circle]{#1};}

\begin{document}
%
\title{Cyber-Physical Energy Systems Modeling,\\Test Specification, and Co-Simulation Based Testing}
%
%
%


\author{\IEEEauthorblockN{
A.~A.~van~der~Meer\IEEEauthorrefmark{1}, 
P.~Palensky\IEEEauthorrefmark{1},
K.~Heussen\IEEEauthorrefmark{2}, 
D.~E.~Morales~Bondy\IEEEauthorrefmark{2},
O.~Gehrke\IEEEauthorrefmark{2}, 
C.~Steinbrink\IEEEauthorrefmark{3},\\ 
M.~Blank\IEEEauthorrefmark{3},
S. Lehnhoff\IEEEauthorrefmark{3},
E.~Widl\IEEEauthorrefmark{4},
C.~Moyo\IEEEauthorrefmark{4}, 
T.~I.~Strasser\IEEEauthorrefmark{4},
V.~H.~Nguyen\IEEEauthorrefmark{5}, 
N.~Akroud\IEEEauthorrefmark{6},
M.~H.~Syed\IEEEauthorrefmark{7},\\
A.~Emhemed\IEEEauthorrefmark{7},
S.~Rohjans\IEEEauthorrefmark{8},
R.~Brandl\IEEEauthorrefmark{9}
A.~M.~Khavari\IEEEauthorrefmark{10}
}
\vspace{0.25cm}
\IEEEauthorblockA{
\IEEEauthorrefmark{1}Delft University of Technology, Delft, The Netherlands, a.a.vandermeer@tudelft.nl}
\IEEEauthorblockA{
\IEEEauthorrefmark{2}Technical University of Denmark, Lyngby, Denmark}
\IEEEauthorblockA{
\IEEEauthorrefmark{3}OFFIS e.V., Oldenburg, Germany}
\IEEEauthorblockA{
\IEEEauthorrefmark{4}AIT Austrian Institute of Technology, Vienna, Austria}
\IEEEauthorblockA{
\IEEEauthorrefmark{5}University Grenoble Alpes, G2Elab, Grenoble, France}
\IEEEauthorblockA{
\IEEEauthorrefmark{6}Ormazabal Corporate Technology, Bilbao, Spain}
\IEEEauthorblockA{
\IEEEauthorrefmark{7}University of Strathclyde, Glasgow, United Kingdom}
\IEEEauthorblockA{
\IEEEauthorrefmark{8}HAW Hamburg University of Applied Sciences, Hamburg, Germany}
\IEEEauthorblockA{
\IEEEauthorrefmark{9}Fraunhofer Institute of Wind Energy and Energy System Technology, Kassel, Germany}
\IEEEauthorblockA{
\IEEEauthorrefmark{10}European Distributed Energy Resources Laboratories (DERlab) e.V., Kassel, Germany}
}

\maketitle

\begin{abstract}

The gradual deployment of intelligent and coordinated devices in the electrical power system needs careful investigation of the interactions between the various domains involved.
Especially due to the coupling between ICT and power systems a holistic approach for testing and validating is required. 
Taking existing (quasi-)~standardised smart grid system and test specification methods as a starting point, we are developing a holistic testing and validation approach that allows a very flexible way of assessing the system level aspects by various types of experiments (including virtual, real, and mixed lab settings).
This paper describes the formal holistic test case specification method and applies it to a particular co-simulation experimental setup. The various building blocks of such a simulation (i.e., FMI, mosaik, domain-specific simulation federates) are covered in more detail. 
The presented method addresses most modeling and specification challenges in cyber-physical energy systems and is extensible for future additions such as uncertainty quantification.

\end{abstract}


%
\IEEEpeerreviewmaketitle

\section{Introduction}
\label{sec:introduction}



The decarbonisation of the European power generation requires a high penetration of distributed, Renewable Energy Sources (RES). Their intermittent behaviour and limited storage capabilities present new challenges to power system operators in maintaining the security of supply and the power quality \cite{EC-Energy2011}. However, advanced Information and Communication Technologies (ICT), automation approaches, and corresponding algorithms provide new possibilities and intelligent solutions for operating power grids in a more optimized way \cite{Palensky2013b, Strasser2015}. As a consequence of these developments the traditional power system is transformed into a Cyber-Physical Energy System (CPES), a smart grid \cite{Farhangi2010}. Previous and ongoing research activities have mainly focused on validating certain aspects of smart grids, but until now there is no integrated approach for analysing and evaluating complex configurations in a cyber-physical systems manner available \cite{Strasser2016}. 

In the process of designing and developing a specific solution, validating and testing the correctness is an essential stage. CPES like smart grids are a combination of different technologies across heterogeneous domains (power, ICT/automation, markets, customer behaviour, etc.), which have mutual interactions and inter-dependencies. Before deploying algorithms and solutions, field tests are needed to evaluate the integration on a system level, addressing all relevant domains. Up-to-now such a cyber-physical approach for designing, analysing, and validating smart grid systems is missing. The existing laboratory-based testing approaches often focus on a certain sub-system (or business sector) and its components. The integration of components---including analysis and evaluation---is not yet addressed sufficiently in a holistic manner. 

Simulation-based experiments are one of the alternative testing approaches that can cover multiple domains \cite{Podmore2010}. However, the development of smart grid solutions and technologies has increased the need for a more integrated simulation approach covering all targeted areas \cite{Mets2014, Palensky2016}. A general framework for smart grid validation and roll-out is necessary. One of the main barriers to this has been the lack of design approaches and corresponding software tools that are capable of simulating power and ICT systems holistically \cite{Strasser2016}.

The lack of such system validation for smart grids is especially addressed by the European ERIGrid project \cite{ERIGridURL}. By providing a Pan-European research infrastructure ERIGrid supports the technology development as well as the roll-out of smart grid solutions and concepts. It tackles a holistic, CPES-based approach by integrating European research centres and institutions with outstanding research infrastructures and jointly develops common methods, concepts, and procedures. 

The aim of this paper is to discuss advanced modeling approaches, a formal specification of corresponding validation scenarios, and co-simulation based testing methods for CPES that are being developed in ERIGrid. 

The rest of this paper is organized as follows: Section~\ref{sec:holistic_test_specification} gives an overview of the proposed methodology for a holistic test description, while Section~\ref{sec:co_simulation} discusses the co-simulation of CPES. Necessary software and component interfaces which are being used in the ERIGrid co-simulation environment are introduced in Section~\ref{sec:software_and_interfaces}. A proof-of-concept example of both the test specification and the co-simulation based experiments is described in Section~\ref{sec:holistic_testing}. Finally, a discussion and an outlook on planned future research is provided in Section~\ref{sec:evaluation}.

%

\section{Methodology for Holistic Test Description}
\label{sec:holistic_test_specification}


Smart grid components and functions are embedded in a distributed and complex infrastructure. The formulation and specification of test requirements is therefore both a technical and conceptual challenge. On the one hand, the technical aspect of carrying out integrated experiments by means of real-time, hardware-in-the-loop or co-simulation technologies has received significant attention in recent years \cite{Palensky2016, Palensky2016b}. On the other hand, significant conceptual development and consolidation has been achieved in the ICT aspects of smart grids through standardization and harmonization of use case descriptions, system architecture, and interoperability test specification \cite{santodomingo2014sgam,heussen2015use,cen2012SGAM,cen2014interop}. 
These technical and conceptual developments have been largely pursued in independent tracks. A wide gap remains in formulating test specifications that facilitate the holistic evaluation of the integrated cyber-physical system. The initial step towards carrying out holistic tests of CPES is to harmonize the concepts and testing methods used by Research Infrastructures (RIs). Further steps then entail the harmonization of the test evaluation methodology, the harmonization of RI internal and external interfaces, as as well as interchangeable configuration descriptions. This will be of particular importance for managing the high complexity of experiments conducted across multiple RIs.

In order to frame this approach we propose the following definition of \textbf{holistic testing}: \emph{``The process and methodology for evaluation of a concrete function, system or component within its relevant operational context with reference to a given test objective.''} 
A challenge is therefore to formalize the complete cyber-physical system context and test criteria in a common framework. A first outline of the holistic testing approach was presented in \cite{Blank2016}, where a focus was placed on the concept of holistic testing and a corresponding methodology. The challenge of addressing multiple RIs with a single test case, as outlined in \cite{Blank2016}, is within the scope of the presented approach, but details will be omitted for brevity.

\subsection{Incremental Test Description}
\begin{figure}[!t]
\centering
\includegraphics[width=0.97\columnwidth]{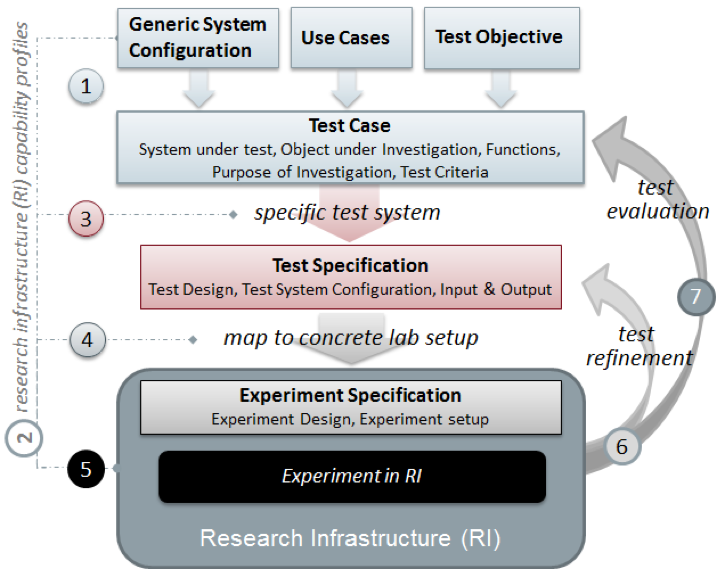}
\caption{The main steps in the ERIGrid methodology for holistic testing.}
\label{fig:holistic}
\end{figure}

Based on these concepts, we have formulated a procedure for holistic test specification, which defines steps for formulating a \emph{concrete} and \emph{holistic} test description (cf. Fig.~\ref{fig:holistic}, steps \tikzCircle{1}-\tikzCircle{4}). In this process, we separate the questions ``what needs to be tested?'' and ``why does it need testing?'', which are answered in step \tikzCircle{1}, from the question ``how should it be tested?'', which is answered in step \tikzCircle{3}. The above questions can be answered independent of the respective testing infrastructure\footnote{The terms lab, research infrastructure (RI), test infrastructure, as well specific co-simulation environments are used interchangeably here.}, so the question  ``what infrastructure is available to carry out the test?'' is addressed to define the available RIs' capabilities in step \tikzCircle{2}. Finally, by answering ``how should the available infrastructure be configured to carry out the specified test in a concrete experiment'' in step \tikzCircle{4}, the experiment specification is completed. 

\begin{table*}[t]
\centering
\caption{Types of System Configurations}
\label{SC-types}
\begin{tabular}{|p{2.5cm}|p{2.2cm}|l|l|p{8cm}|}
\cline{1-5}
\textbf{Name / Purpose}&\textbf{Context} & \textbf{Generic/Specific} & \textbf{SCType} & \textbf{Explanation}  \\ \cline{1-5}
Function-System Alignment &Use Case  & GSC & UC-GSC         & As SGAM domains \& zones: reference designation for functions, independent of test case.  \\ \cline{1-5}
Test Case context model & Test Case (step \tikzCircle{1})  & GSC  & TC-GSC & Establishes type conventions for test case: relevant SC component types, domains, etc., and categorically identifies the SuT.      \\ \cline{1-5}
Test System & Test Specification (step \tikzCircle{3}) & (S)SC& TS-SC & A concrete instance of TC-GSC to address a specific OuI and specific set of test criteria. \\ \cline{1-5}
Experiment Setup & Experiment Specification (step \tikzCircle{4})& (S)SC & E-SC & The configuration and interconnection of RI components, representing the SuT, and including OuI. \\ \cline{1-5}
RI description & RI database  (step \tikzCircle{2})& (S)SC     & RI-SC    & Configuration and components of an RI, including \emph{potential} multiplicity and  connectivity of RI components. \\ \cline{1-5}
RI information model& RI profiling & GSC & RI-GSC   & RI profiling information model, here specific to CPES laboratories.  \\ \cline{1-5}
\end{tabular}
\end{table*}  
   
In order to formulate the holistic test case, three inputs are required:
\begin{itemize}
\item A \emph{Generic System Configuration}, which is a description of the overall system configuration (and assumptions) within which we seek to test an object (or subsystem).
\item A set of \emph{Use Cases} that describe the sequence of actions/functions that are expected of the tested object.
\item The \emph{Test Objective}, which is the purpose for carrying out the tests, stating the overall evaluation objective.
\end{itemize}

\noindent Based on the previous inputs, the \textbf{test case} describes the following concepts:
\begin{itemize}
\item A \emph{System under Test} (SuT) that identifies the system boundaries of an abstract test system entailing all relevant interactions requiring investigation.
\item The \emph{Object(s) under Investigation} (OuI) identifying the systems, subsystems, or components within the SuT to which the test criteria will be applied.
\item The \emph{Domain(s) under Investigation} (DuI), which identifies the relevant physical or cyber-domains (and sub-domains) which are of interest for the test parameters and connectivity.
\item With respect to the use cases, \emph{Function(s) under Test} (FuT) and \emph{Function(s) under Investigation} (FuI), which describe the relevant operations relevant to the SuT (in the FuT) and the ones that are being investigated with respect to the OuI (in the FuI).
\item The \emph{Purpose of Investigation} (PoI), which details the test objective and sub-objectives by qualifying each objective as either \emph{validation}, \emph{verification}, or \emph{characterization}.
\item \emph{Test Criteria}, which state the metrics that need to be evaluated for each of the objectives formulated in the PoI.
\end{itemize}

\noindent With these concepts, we are able to define a \textbf{test specification}, which consists of:
\begin{itemize}
\item A \emph{test system configuration} which defines how the OuI is going to be embedded in the SuT.
\item The \emph{input/output} parameters of the system which will be varied, observed, and evaluated.
\item The \emph{test design}, which defines the manner in which the test will be carried out.
\end{itemize}

Finally, given a test specification, we can map the testing requirements with the capabilities of one or more RI(s) in order to carry out the experiment. This mapping (step \tikzCircle{4} in Fig.~\ref{fig:holistic}) leads to a concrete \textbf{experiment specification}, which identifies the laboratory components and devices that reflect the required functions, domains and connections specified in the previous steps in order to execute the test.

The experiment specification is the last step of the description phase before an experiment can be conducted (step \tikzCircle{5} in Fig.~\ref{fig:holistic}). After the experiment, a pre-assessment of the results is required (step \tikzCircle{6}) to decide whether the test specification will have to be adapted for a re-run of the experiment in a modified setting, or if the results are suitable for a final evaluation of the experiment (test evaluation, step \tikzCircle{7}).

The process outlined above is equally applicable to tests involving multiple RIs. Examples of multi-RI  testing include the mapping of identical test specifications to different RIs or the synchronous execution of one large experiment spanning multiple communicating RIs. However, each of these cases introduces specific additional requirements such as ensuring the comparability of results obtained on dissimilar laboratory hardware, or the need to specify the real-time interaction at the inter-RI level. These needs are being addressed by the ERIGrid project but are beyond the scope of this paper.

\subsection{System Configurations (SC)}

The specification of system configurations is central to the test description, similar to other related specification work (e.g., smart grid use cases with reference to the Smart Grid Architecture Model (SGAM), information modeling for power system ICT via the Common Information Model (CIM), or other applications of the Unified Modeling Language (UML) or the Systems Modeling Language (SysML)). To integrate the SC description methods used in smart grid disciplines (e.g., electrical, ICT, or thermal systems), ERIGrid adopted and generalized the basic system description concepts that are employed in the power systems CI). To avoid overly complex specification details, the following upper ontology model has been identified: \textit{System}s are composed of \textit{Components} and are themselves components.  Components have \textit{Terminals}, which may have a directionality and are associated with one \textit{Domain}. Domains can be structured hierarchically. Two or more terminals associated with the same domain can be connected using a \textit{Connection Point}. All of the above are \textit{System Configuration Objects}, and a set of them composes a \textit{System Configuration Container}, which has a system configuration type (\textit{SCType}) attribute. \textit{Constraints} can be associated with any type of system configuration object. 
Table~\ref{SC-types} lists the six relevant types of system configurations. 

A domain-independent graphical realization of an SC description (TS-SC) is found in Fig.~\ref{fig:UC-GSC-TC1}, where a test system is specified. 
Using another graphical convention, Fig.~\ref{fig:TC1exp}, defines a co-simulation Experiment SC (E-SC). 
When formulating the E-SC based on TS-SC, interactions among SuT elements are fully represented, whereas non-SuT components (non-grayed background in Fig.~\ref{fig:UC-GSC-TC1}), are represented by simple equivalents. 

While the SC abstraction is useful, the number of SC variants listed in Table~\ref{SC-types} may be confusing. Two main aspects of distinction should be noted: 
Firstly, there are \textit{RI-oriented system configurations} (RI-GSC, RI-SC,E-SC) and \textit{real-world oriented system configurations} (UC-GSC, TC-GSC, TS-SC). The former describe the RI capabilities and concrete RI objects that are eventually present in an experiment. This may include a physical grid or amplifier in a lab, but also simulation facilities. The latter types of SC aim to represent aspects of the real-world. 
Secondly, there are \textit{generic} and \textit{specific} system configurations. The generic SCs (UC-GSC, TC-GSC,RI-GSC) define types or classes of objects, whereas the specific SCs  (TS-SC, RI-SC, E-SC) define concrete instances of SC elements.








\section{Co-simulation of CPES using Mosaik and FMI}
\label{sec:co_simulation}



The software co-simulation in ERIGrid is realized by employing the framework mosaik as a co-simulation master and interfacing all simulators via the Functional Mock-up Interface (FMI) standard.
In the following, the components of this co-simulation platform are briefly introduced.
Subsequently, the coupling between mosaik and FMI is elucidated, as well as the utilization of the platform in the project.

\subsection{Framework and Interfaces}

\subsubsection{Mosaik}
Mosaik is a co-simulation framework that is designed for easy integration of simulators and flexible creation of co-simulation scenarios.
These design goals are achieved via two Application Programming Interfaces (API).
The Component-API allows integration of simulators via description of the accessible data and a set of interface functions.
The API has been ported to different programming languages to provide a broad user support.
The Scenario-API provides a set of functions to establish data flow connections between simulators and execute the co-simulation.
The execution is managed by a discretely timed scheduling algorithm.
Two versions of mosaik have been developed so far.
Mosaik~1 \cite{Schutte2013} incorporates a domain specific language for the two API types with the goal of automatic consistency checking of co-simulation scenarios.
It ultimately proved to be too inflexible for practical work.
Mosaik~2 (e.g. \cite{Buscher2014b, Lehnhoff2015}) is an enhanced, streamlined version of the software that is purely based on Python and provides more concise and flexible APIs.
\subsubsection{Functional Mock-up Interface}
FMI \cite{MODELISARConsortium2014} is a standard for the interfacing of simulators and simulation models that is supported by various modeling and co-simulation tools.
It is comparable with the mosaik Component-API in the sense that it provides a (XML-based) data model and a set of interface functions.
However, FMI is more complex and powerful than its mosaik counterpart.
The reason for this lies in the fact that the Component-API is specifically geared to the mosaik scheduling algorithm while FMI has been designed for interaction with a variety of master algorithms.
The design of the master is not part of the FMI standard so that the complexity of FMI is the price of its flexibility.
A simulator that employs the FMI standard is capsuled in a Functional Mock-up Unit (FMU).
There are two types of FMI so that FMUs may possess different structures.
FMI for Co-Simulation (FMI-CS) assumes that the FMU includes a solver and thus may independently simulate when called.
FMI for Model Exchange (FMI-ME), on the other hand, expects the master algorithm to solve the simulation model provided by the FMU.

To facilitate the handling of FMUs, the FMI++ toolbox has been developed \cite{Widl2013}.
It provides a more high-level set of interface functions that is still usable by a variety of master algorithms.
Furthermore, it includes a number of utility methods, e.g., for rollback and interpolation, as well as a set of readily usable integrators to solve FMI-ME systems.

\subsection{Coupling between mosaik and FMI}
Mosaik is used as the co-simulation framework in ERIGrid due its good usability while providing robust scheduling at the same time.
For the simulator interfacing, however, FMI is employed due to its wide acceptance and usefulness for potential future exchange of schedulers.
Therefore, a mapping between the Component-API and FMI has to be established to grant mosaik access to simulators encapsulated in FMUs.
Such a mapping has already been conducted successfully by \cite{Rohjans2014} via the use of the FMI++ toolbox.
However, their approach employs mosaik~1 and an outdated version of FMI++.
Furthermore, only FMI-ME is supported whereas ERIGrid will utilize both types of FMI.
As a consequence, coupling is suggested that employ the more streamlined interface functions of mosaik~2 and the new FMI++ version.
Separate interfaces are established for FMI-ME and FMI-CS.
Each of them is kept as generic as possible so that integration of several FMUs of the same type requires only an adjustment of parameter sets.

As shown in \cite{Rohjans2014}, a mapping has to be established between the mosaik API functions and the FMI++ functions.
ERIGrid employs a mapping similar to the one presented in that work, although using updated function references.
Similarly, the XML-based description of model variables has to be mapped onto the mosaik model description.
FMI includes a complex variable description with data fields like ``variability'' and ``causality'', many of which do not possess a counterpart in mosaik.
Parameter-type variables are also parameters in mosaik. 
Variables with the causality ``input'' or ``output'' are summarized as ``attributes''.
Nevertheless, the exact causalities as well as the data types (real, integer, etc.) have to be stored within the FMI-mosaik interface in order to select the correct getter and setter functions provided by FMI++.

\begin{figure}[!t]
\centering
\includegraphics[width=0.65\columnwidth]{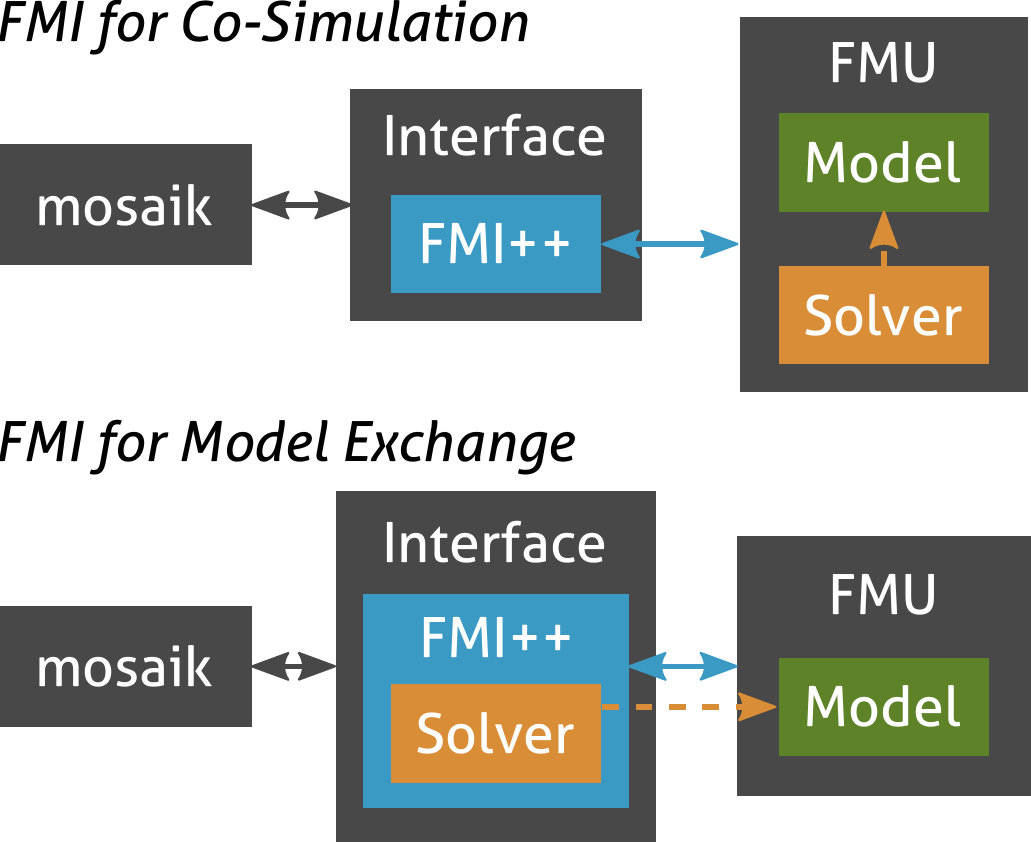}
\caption{Interfacing of mosaik for the different FMI types.}
\label{fig:interf}
\end{figure}

As mentioned before, FMI-ME expects the master algorithm to solve the simulation model.
This notion generally conflicts with the co-simulation concept, e.g., mosaik does not include any solvers.
This problem is circumvented by employing the integrators provided by FMI++ within the FMI-mosaik interface.
Thus, whenever mosaik calls for the FMU to execute a step, the interface solves the model for this time step.
The desired integrator type can be set, along with other technical details like the time step size, as a ``simulator parameter''.
FMI-ME is, in summary, supported via a form of ``co-simulation via capsuling'' as illustrated in Fig.~\ref{fig:interf}.

\subsection{Application in ERIGrid}
Mosaik fullfills two roles in ERIGrid.
On the one hand, it acts as an environment for exemplary co-simulation test cases, serving its usual purpose.
This setup utilizes the coupling between mosaik and FMI-CS to allow integration of complex tools like \textsc{PowerFactory}.
On the other hand, mosaik is also used as a testbed for a library of newly developed smart grid component models.
These models are supplied following the FMI-ME standard to enable versatile future use.

The developed models should be capable of being integrated into a variety of different simulation environments like MATLAB or \textsc{OpenModelica}.
Therefore, FMI-ME is the preferred standard since it provides more freedom in the model usage.
Due to the interface presented above, mosaik is still able to integrate these FMUs and serve as an environment for integration testing.
For testing, an FMU is integrated and presented with pre-defined input data from a generic data source component.
The output may be stored in a database component and analyzed for unexpected behavior.
As a matter of fact, mosaik's flexible Scenario-API allows a convenient way to automatically test sensitivities towards different parameter values, temporal resolutions, integrator types, and so forth.
Mosaik's capabilities for such a type of black box analysis have already been demonstrated in \cite{Lehnhoff2015}.

\section{Software and Component Interfaces\\applied in ERIGrid}
\label{sec:software_and_interfaces}

%
%

Challenges in transferring R\&D results into real-world applications often arise already in the design phase, due to incompatible simulation-based approaches adopted by researchers and industry that prevent a consolidated solution for validation.
Hence, one goal of the ERIGrid project is the demonstration of the feasibility of FMI-based co-simulation and model exchange for validating Smart Grid applications, in order to promote tool and model interoperability and stress the importance of this topic in the FMI community.
To this end, the proof-of-concept studies performed in ERIGrid need to be representative and demonstrate the added value of such an approach.
In the following, the rationale behind the choice of tools and models in this context is explained.

\subsection{Simulation domains}
\label{sec:simulation_domains}

Modeling and simulation of smart grid scenarios comprise aspects from various technical domains, with a broad variety of available simulation tools for each of these domains.
For the purpose of ERIGrid, three particular domains have been identified:
\begin{enumerate}

  \item \textit{Power systems}:
  The technical infrastructure for the distribution of electricity is obviously the core element for smart grid applications.
  For ERIGrid, \textsc{PowerFactory}\footnote{See \texttt{\url{http://www.digsilent.com/}}} will be primarily used to simulate power systems.
  
  \item \textit{Communication}:
  Virtually any smart grid application relies in one way or the other on information exchange.
  The discrete-event network simulator \textsc{ns-3}\footnote{See \texttt{\url{https://www.nsnam.org/}}} has been selected as main tool for ERIGrid to simulate dedicated and general-purpose ICT systems.
  
  \item \textit{Automation and control}:
  The potential complexity of controllers used in smart grid applications necessitates dedicated tools for their implementation.
  For ERIGrid, \textsc{MATLAB/Simulink}\footnote{See \texttt{\url{http://www.mathworks.com}}} will be primarily used to provide the needed controller functionality for simulations.

\end{enumerate}

Furthermore, a dedicated library of models will be developed using \textsc{OpenModelica} and \textsc{MATLAB/Simulink}, which will complement and expand the functionality provided by the domain-specific tools.
The selection of the tools and models was driven by several factors, such as previous experience of partners, availability of FMI-compliant interfaces or details regarding licenses.

\subsection{Tool compatibility}
\label{sec:tool_compatibility}

Each of the selected tools represents the state-of-the-art in their respective domain.
As such, this selection provides a representative case that can serve as a relevant proof-of-concept for smart grid co-simulation.
However, the selected tools implement very contrasting modeling and simulation paradigms and to prove the capability and point out the shortcomings of FMI-compliant interfaces to successfully handle this heterogeneous combination is a key issue.

\subsubsection*{Power system modeling}
Tools for this domain rely on continuous time-based modeling paradigms, typically representing individual components by (sets of) differential algebraic equations.
They enable the simulation of the evolution of the system state either with the help of models that depend explicitly on time (RMS and EMT simulation) or by computing a series of subsequent power flow calculations (quasi-static simulation).
Even though \textsc{PowerFactory} does not offer the functionality to export models in a way that is compliant with FMI-ME, it provides an API that allows to interact with it~\cite{Stifter2013}.
The functionality of this API has been successfully mapped to the that specified for FMI-CS\footnote{See \texttt{\url{http://powerfactory-fmu.sourceforge.net}}}.

\subsubsection*{Communication network modeling}
Simulators for this domain use abstractions of the deployed hardware and software that allows the representation of communication processes as a sequential processing and transmission of (virtual) messages and signals. 
Hence, communication network simulators commonly implement discrete event-based simulation paradigms, where each event marks a significant step of message processing or transmission.
The network simulator \textsc{ns-3} has already been successfully used for co-simulation, see for instance~\cite{Ciraci2014}, but no FMI-compliant interface is available so far.
Based on previous theoretical work dealing with the capability of FMI to handle discrete event semantics~\cite{Broman2013, Tripakis2014}, an interface according to the FMI-CS specification is being developed in ERIGrid for \textsc{ns-3}.

\subsubsection*{Automation and control}
For implementing control algorithms \textsc{MATLAB/Simulink} has been chosen, due to its versatility as well as its popularity and widespread use for this purpose.
The capability of FMI to encapsulate the functionality of discrete event-driven controller models has been already discussed in~\cite{Widl2015}.
\textsc{Simulink} implements a continuous time-driven paradigm and allows to translate models into FMUs for ME (discussed next in Section~\ref{sec:model_library}).
\textsc{MATLAB} on the other hand is a multi-purpose computational environment, which adheres to no specific modeling paradigm and provides by itself no notion of time.
However, it is possible to put \textsc{MATLAB}'s full functionality at the disposal of the user via an approach that is compatible with FMI-CS\footnote{See \texttt{\url{http://matlab-fmu.sourceforge.net}}}.

\subsection{Model library}
\label{sec:model_library}

The model library developed for ERIGrid aims to facilitate and accelerate the integration and validation of smart grid solutions by extending the functionality of available tools with specific models.
To this end, the models will be compliant to the FMI-ME specification, in order to ensure a tool-independent implementation that improves reusability.
However, even though FMI-ME is a well accepted specification, basically no present-day proprietary power systems simulation tool has incorporated the FMI-ME standard so far.
By demonstrating its use and potential benefits within ERIGrid, it is hoped that the adoption of FMI-ME within the power industry is further propelled.

In the following, examples of models being developed for each of the considered simulation domains and their potential use to the wider community are presented.
All the models will be developed either in \textsc{Matlab/Simulink} or \textsc{OpenModelica}, which allow to export models as FMUs for ME.

\subsubsection*{Power system components}

A number of models specific to power systems are currencly under development for FMI-ME, including a reduced-order distribution dynamic equivalent model, FlexHouse thermal model, PV and battery models, and aggregated wind turbine dynamic model. This set of tool-independent models can potentially be used for different simulation studies and proof of concepts.

For example, the reduced-order dynamic equivalent model represents an actual \SI{115}{kVA} smart grid laboratory at the University of Strathclyde. Although simulation tools such as \textsc{PowerFactory} provide many component models, this reduced-order equivalent model will allow for the dynamics of a real smart grid laboratory to be incorporated in simulation studies. This enables large-scale system studies to be undertaken with the least amount of computational requirements.

\subsubsection*{Communication network modeling}
The difficulty of incorporating the effects of communication within a power system simulation tool (due to its discrete event-based nature) often leads to simplistic fixed-time delay models being utilized.
However, within ERIGrid, a representative communications model that can be easily utilized within a power system simulation is under development. It will rely on various communication network parameters as inputs and will calculate delays based on stochastic equations taking into consideration a set of defined uncertainties ~\cite{Dambrauskas2017}.
By employing the IEC~61850~GOOSE protocol, the model will delay a signal based on a value chosen by means of a Gaussian distribution.
This will enable power system studies to incorporate a simple yet more realistic, representative delay.   

\subsubsection*{Automation and control}
A large number of smart grid applications increasingly depend upon reliable measurements being obtained from within the network. Recently, network critical applications, such as protection, are dependent upon utilizing accurate measurements from a large number of \emph{Phasor Measurement Units}~(PMU).
Within ERIGrid, both P-Class and M-Class PMU models compliant with IEEE C37.118.1a are being developed as in~\cite{Roscoe2013}.
These models can be utilized by the wider community for development of novel applications that rely on data from PMUs.


\section{Holistic Testing: Proof-of-Concept by Co-simulation}
\label{sec:holistic_testing}

\subsection{Modelling and Simulation needs}


   
Generally speaking, software experiments can be subdivided into
monolithic software experiments (one domain, one tool), multi-domain
tools (multiple domains, one tool), hybrid models (multiple tools, one
domain), and heterogeneous modeling (i.e., multiple domains and
tools). The latter is predominantly realized with co-simulation and
will be the main focus of the proof-of-concept of the above proposed holistic test
case formalization \cite{Palensky2016, Palensky2016b}.

The application of the mosaik framework and the FMI opens up a massive spectrum of co-simulation possibilities,
but gives also rise to development and implementation challenges. It
has been decided to define three exemplary test cases that at least
contain one co-simulation experiment, each of which treats one
specific development need:

\begin{enumerate}
\item interfacing software tools with physical controllers
\item signal-based synchronization
\item cyclic dependencies between simulation federates
\end{enumerate}

\noindent Software/hardware interfacing will be studied using an on-load tap changer and its control. Both can be implemented as hardware or emulated while synchronizing in wall clock time. Signal-based synchronization is addressed by an on-load tap changer that is regulated by a distributed voltage controller, the ICT aspects of which play a prominent part in the overall behaviour of the SuT.    

In the forthcoming we will focus on the so-called cyclic dependency issue that arises within mosaik. Coupling continuous simulators necessitates
synchronization of interface variables during all stages of
execution. By default, however, mosaik only allows the specification of one-way
dependencies between simulation federates and defining cyclic
dependencies is non-trivial. To test this particular feature, a rather
simple simulation setup that still allows studying the problem would
be optimal. For the sake of realism we use a power system model, being
simulated in tool 1, and an aggregated wind power plant model, which
is simulated in tool 2. This test bed potentially allows the inclusion
of controls and phenomena with diverse time constants (e.g.,
supervisory frequency control, voltage control, fault-ride through)
and of different nature (i.e., discrete versus continuous).




\subsection{Test Case (step \textcircled{1} in Fig.~\ref{fig:holistic})}

\subsubsection{Use Case and Function under Test}
The Use Case treated here is the capability of a Wind Power Plant (WPP) to stay connected during a three phase short circuit inside the transmission system. This capability is referred to as Fault Ride-Through (FRT) and is often laid down in grid codes, which require conformance at the coupling point (i.e., the legal boundary between the power park module owner and the transmission system operator)\cite{EUGridCode}. Hence, all individual wind turbines need to jointly fulfill this common requirement. Aside from FRT the WPP needs to comply to a myriad of other regulations. Essential for FRT because of the mutual interaction are reactive power and voltage control. Thus, the FuT are 1) FRT capability and 2) reactive power control. The PoI is to \emph{validate} the compliance of the WPP as a whole to the voltage against time profile during voltage sags and against a voltage-reactive power curve for normal operating conditions.

\subsubsection{System Configuration and SuT}

\begin{figure}[!t]
\centering
\includegraphics[width=0.88\columnwidth]{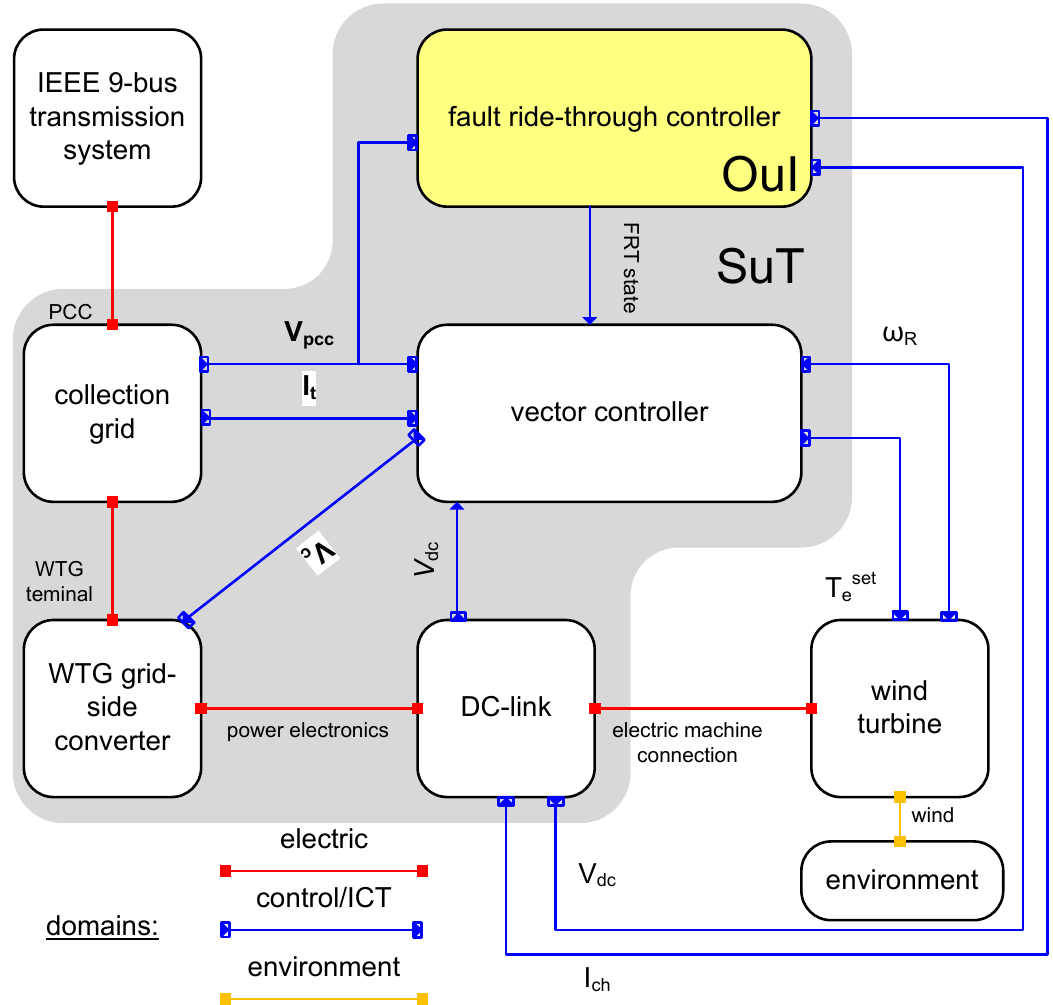}
\caption{\label{fig:UC-GSC-TC1}Test system configuration (TS-SC) of the example holistic test case.}
\end{figure}

The test system SC (see Table~\ref{SC-types} for the terminology) and the SuT of this test case are
shown in Fig.~\ref{fig:UC-GSC-TC1}. It consists of the transmission
system, the collection grid of the WPP, the individual
Wind Turbine Generator (WTG) and its controls. The connections between the
components and subsystems reside in a domain, in this case electric, control, or environment. 

The SuT is the part of the system configuration
comprising the FuT.  Although the actual
implementations of FRT compliance are done inside the individual wind
turbine controllers, FRT verification is required at the coupling
point. Hence the boundary of the SuT is between the transmission
system and the WPP collection grid. By means of vector control the protection circuits in the DC-link dynamically separate the electromechanic part of the WTG from the grid interface. This
allows us to not take the wind turbine itself (i.e., aerodynamic
conversion, drive train, electromechanic conversion, stator-side
controls, pitching, etc.) into consideration. This also bounds the DuI to the electric and control/ICT domains. 

As for verification of voltage and reactive power control the
interactions are foreseen to mainly occur in the collector grid. A
supervisory WPP controller usually sets reference values for either
voltage, reactive power, or power factor, and the individual WTGs need to track these set points locally. Interaction with FRT
is foreseen in case additional reactive current injection is engaged
during faults. The SuT for this FuT hence comprises the WPP collection grid, the grid-side converter, and its vector controls.

\subsubsection{Test Criteria and FuI}

The set of system configuration, use cases, and functions under test cater for the evaluation of two specific test criteria, namely

\begin{enumerate}
\item The validation of the ability of the WPP to maintain synchronism with the external system during voltage dips, as defined in a voltage against time curve; 
\item The validation of the conformance of the WPP to track a reactive power against voltage curve by means of a centralised voltage control mechanism.
\end{enumerate}

\noindent To test the former, both functions under test are significant as reactive power control during faults may boost the voltage magnitude, alleviating the FRT duty of the WPP as a whole. Hence, the FuI are the same as the FuT for the first test criterion. For the second, the FRT mechanism is superfluous to consider as the criterion is merely valid for normal operating conditions. Thus, the FuI is the centralised voltage control of the WPP.

\subsection{Test Specification (step \textcircled{3} in Fig.~\ref{fig:holistic})}


For the sake of brevity, we limit the test specification to the FRT compliance and assume a similar procedure for evaluating the test criterion concerning the reactive power at the coupling point. The FuI here is the set of actions the FRT controller has to take to fulfill the test criterion. Hence, the OuI is the FRT controller.  At this stage we need to further quantify the test criterion, and need to define a more specific system configuration.  

\subsubsection{Specific Test Criterion and Test System Configuration}
\begin{figure}[!t]
\centering
\includegraphics[width=0.75\columnwidth]{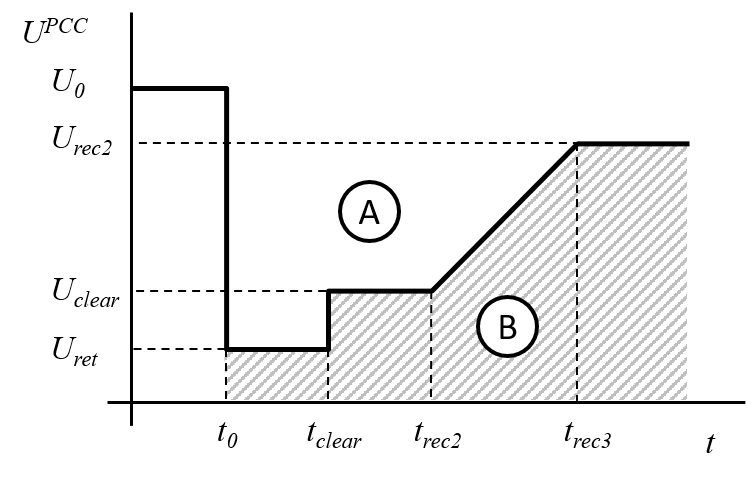}
\caption{\label{fig:FRT} Fault ride through voltage profile.}
\end{figure}

The test criterion is refined according to Fig.~\ref{fig:FRT}. It shows the time versus positive-sequence voltage profile that the WPP needs to be able to endure. In case the point of common coupling (PCC) voltage enters the grey area (i.e., $U^\textrm{PCC}<U_\textrm{ret}$,  \tikzCircle{B}) the WPP is allowed to disconnect from the grid. In zone \tikzCircle{A}, the FRT controller must engage internal component protection and while is shall prevent entering zone \tikzCircle{B}. The protection and additional control measures that need to be taken are implemented as a finite state machine, hence a discrete controller \cite{AAvdM2015b}. 
The corresponding test system comprises the IEEE 9-bus transmission system, the SuT and the WTG. The SuT and its internal connections are indicated in Fig.~\ref{fig:UC-GSC-TC1} by the gray area. The IEEE 9-bus benchmark system (i.e., \cite{Anderson2003}), has been adapted to \SI{50}{Hz} and to contain more realistic dynamic behaviour of loads. This TS-SC represents the WPP aggregatedly by a single permanent-magnet, direct drive WTG. Detailed single-line representations of the collection grid and transmission grid have been omitted for brevity.

\subsubsection{Overall design of the test}

The test criterion to verify consists of two parts, a deep-dip part between $t_0$  and $t_\textrm{clear}$ and a recovery part between $t_\textrm{clear}$ and $t_\textrm{rec3}$. The deep-dip part is commonly caused by severe voltage dips that are quickly isolated by protection. The recovery part serves two situations: 1) the voltage amplitude dynamics after fault clearance at or before $t_\textrm{clear}$, 2) the voltage response after a shallow voltage dip cleared after $t_\textrm{clear}$ and does not violate $U^\textrm{PCC}<U_\textrm{clear}$. 

These situations can be replicated by causing a voltage dip at the PCC by means of a self-extinguishing 3-phase-to-ground fault in the external grid. The testing procedure is as follows:
\begin{enumerate}
   \item Determine operating point;
   \item Set short circuit location to $x$ in the IEEE 9-bus system, corresponding dip depth; $U_\textrm{ret}^x$ 
   \item Initiate short circuit at $t_0=0.1\text{s}$;
   \item Clear fault at $t_\textrm{clear}=y$;
   \item Assess test criteria; and
   \item Vary $x$ and $y$ and repeat above sequence 2-5 times.
\end{enumerate}

\noindent The experiment is considered successful if the WPP maintains synchronism under the circumstances defined above. 

\subsection{Experiment Specification (step \textcircled{3} in Fig.~\ref{fig:holistic})}

The test criterion can be evaluated by several experiments. As field testing or pure hardware testing is unfeasible in this case---it will be challenging to convince DSOs and WPP owners to release their system as a testbed---the emphasis will be on pure software and (C)HIL experiments. Example experiment realizations include:

\begin{itemize}
\item Monolithic simulation using MATLAB/Simulink (Simscape Power Systems Toolbox)
\item CHIL in which the grid is simulated by a real time simulator and one of the wind turbine controllers is implemented into a PLC and is connected via I/O
\item Co-simulation between PowerFactory and OpenModelica
\item Co-simulation between PowerFactory and mosaik using both FMI for CS and ME
\end{itemize}

\begin{figure}[!t]
\centering
\includegraphics[width=0.75\columnwidth]{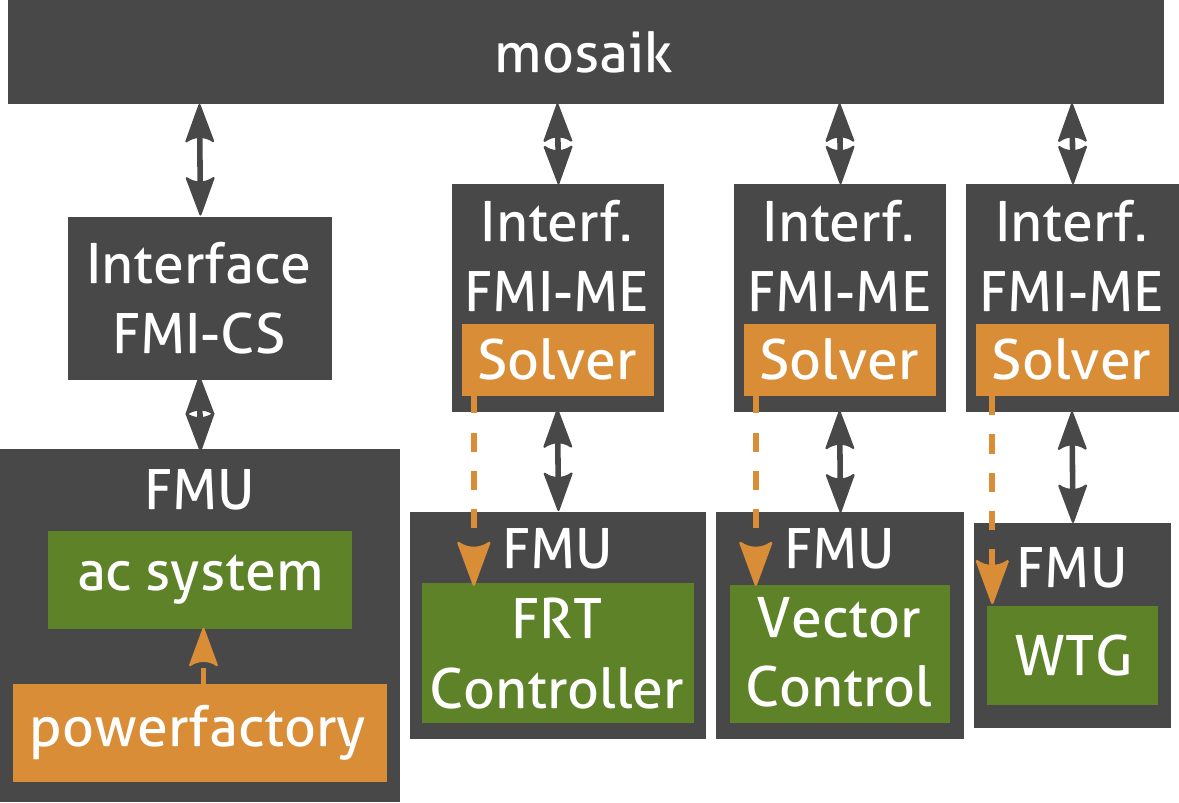}
\caption{\label{fig:TC1exp} Experimental implementation of TC1 in terms of mosaik and FMI.}
\end{figure}

\noindent The latter will now be explained in more detail as it combines all the previously introduced building blocks (i.e., FMI++, mosaik, simulation tools). Fig.~\ref{fig:TC1exp} shows the distribution of the system configuration across \textsc{PowerFactory} and FMUs for ME. The ac grid part of the test system is simulated in \textsc{PowerFactory} and comprises the IEEE 9-bus system, the collection grid, and the ac grid interface of the WTG. The latter is modelled by a variable quasi-stationary current source. The set point values of this current source typically come from the master simulator, which is done through the FMI-CS interface. This wrapper connects to mosaik, which on its turn orchestrates the overall simulation procedure. The wind turbine, its controls, and the FRT controller are tested and implemented in \textsc{Simulink} and subsequently exported as FMUs for ME. The FMI++ wrapper accounts for the solution routine needed for these FMUs.

This experimental setup combines all the merits that come along with mosaik and FMI++: it allows design and validation of models in a multi-domain simulation environment, couple it to a master simulator by FMI-ME, and a flexible mock-up with a commercial, domain-specific simulation tool by FMI-CS.

\section{Evaluation, Discussion, and Outlook}
\label{sec:evaluation}


This paper discussed a formal approach for holistic test description, which is an invaluable tool for evaluation of the integrated cyber-physical systems. This approach aims to harmonize the concepts and testing methods used by laboratories and research infrastructures to analyze CPES, which are a combination of different technologies across heterogeneous domains (power, ICT/ automation, markets, etc.). In this paper we provided co-simulation by the open-source mosaik framework using the FMI standard as one particular experimental implementation of the holistic test description approach. The extension of mosaik with FMI for co-simulation and model exchange offers a versatile tool chain that allows multi-domain co-simulation, testing, and validation, for components and systems alike. 

As a proof of concept of holistic CPES assessment by co-simulation, this paper discussed the fault ride-through of a wind power plant that is grid connected at sub-transmission level. This test case highlights the benefits of the formal specification method. On use case level it clearly distinguishes the interacting functions under test, on test specification level the behaviour and modeling assumptions can be projected on the test criteria, and the experimental specification allows a very flexible way of assessing the separate sub-systems on diverse experiment platforms.
Next steps include the test evaluation criteria, test refinement, and optimally mapping the research infrastructure to the experiment specification. Future work will include the application of uncertainty quantification (UQ) methods to the test specification and evaluation process, including UQ annotations in the model library, which will enable UQ of the associated co-simulation experiments \cite{Steinbrink2016b}.


The tools and models chosen and implemented in ERIGrid provide a representative example of FMI-compliant state-of-the-art concepts for modeling and simulation of CPES. This will pave the way for improving interoperability and repeatability for simulation-based evaluation and validation approaches, notably by demonstrating the feasibility and illustrating the advantages and shortcomings of using FMI-compliant tools and models. 

\section*{Acknowledgment}

This work is supported the European Community's Horizon 2020 Program (H2020/2014-2020) under project ``ERIGrid'' (Grant Agreement No. 654113). 

\ifCLASSOPTIONcaptionsoff
  \newpage
\fi



\bibliographystyle{IEEEtran}

\bibliography{literature}
%

%






\end{document}